# Exploration in Algorithm Engineering: Modeling Algorithms

**Sabah Al-Fedaghi**

*sabah.alfedaghi@ku.edu.kw*

Computer Engineering Department, Kuwait University, Kuwait

**Summary**

According to some algorithmicists, algorithmics traditionally uses *algorithm theory,* which stems from mathematics. The growing need for innovative algorithms has caused increasing gaps between theory and practice. Originally, this motivated the development of algorithm engineering, which is viewed as experimental techniques related to software engineering. Currently, algorithm engineering is a methodology for algorithmic research that combines theory with implementation and experimentation in order to produce better algorithms with high practical impact. Still, researchers have questioned whether the notion of algorithms can be defined in a fully generable way and discussed what kinds of entities algorithms actually are. They have also struggled to maintain a view that formulates algorithms mathematically (e.g., Turing machines and finite-state machines [FSMs]) while adapting a more applied view. Answering the question of what algorithms have practical applications in software specifications in particular, this paper proposes a diagrammatical definition of an algorithm based on a new modeling machine called a thinging machine (TM). The machine has five actions (e.g., create, process, release, transfer, and receive) that can form a network of machines. The paper explores the application of the definition in Turing machines and FSMs. The results point to the fact that the proposed definition can serve as a middle-ground representation of algorithms, a definition which is between formal specification and the commonly used informal definition (e.g., set of instructions).

## 1. Introduction

According to Sanders et al. [1], the algorithms that drive search engines and the pattern-matching algorithms crucial for reading the human genome are only a few spectacular examples of how algorithms can change our lives. Traditionally, algorithms have been studied from the perspective of mathematical algorithm theory, where the study of algorithms is dominated by mathematical (worst-case) analysis. The growing need for innovative algorithms has caused the gap between theory and practice to increase. Real-world inputs are often far from the worst-case scenarios assumed in theoretical analysis. In extreme cases, promising algorithmic approaches are neglected because a mathematical analysis would be difficult [1]. Originally, this drove the development of algorithm engineering, which is viewed as experimental techniques related to software engineering. Currently, algorithm engineering is a methodology for algorithmic research that combines theory with implementation and experimentation in order to produce better algorithms with high practical impact [2–3].

It is important to understand what algorithms have practical applications in software specification and model-based testing in particular, as well as theoretical applications in areas such as semantics of software or algorithmic completeness of computation models [4]. Researchers (e.g., Gurevich [4]) have questioned whether the notion of an algorithm can be rigorously defined in full generality and discussed the kinds of entities that algorithms actually are. According to Rapaport [5], "As hard as it is to define 'computer,' the notion of 'algorithm' is even murkier, despite the accomplishments of Church, Godel, Kleene, Markov, Turing, Post, and others." Students in computing have struggled to keep the abstract view that formulates algorithms mathematically, as well as the more applied view [6].

According to Gurevich [4], many types of algorithms have been introduced (e.g., classical sequential algorithms, parallel, interactive, distributed, real-time, analog, hybrid, quantum). Algorithms are represented in many models of computation and can be defined as any sequence of operations that can be simulated by a Turing (-complete) machine. Models of computation include Turing machines and finite-state machines (FSMs and are used to analyze various relationships between algorithms, to prove theorems about the computational complexity of algorithms, and so forth [7]. A model of computation is a description of a computation in abstract terms. Computation refers to the transformation of input information into output information [7].

Turing machines are considered one of the foundational models of computability [8]. Turing machines reduce the description of algorithm activity to its simplest elements and can be used to describe an algorithm independent of the architecture of a particular machine. A Turing machine is a machine capable of incorporating a finite set of state machines. It is supplied with an infinite one-way, one-dimensional tape divided into squares that are each capable of carrying exactly one symbol. At any given moment, the machine scans the content of one square, and the *behavior* of the machine is determined by the current state and symbol being scanned [8].

A FSM is a conceptual model that can be used to describe how many things work [9]. FSMs are a type of restricted Turing machine. According to Lee and Seshia [10], a state machine is a model of a system with "discrete dynamics that at each reaction maps valuations of the inputs to valuations of the outputs, where the map may depend on its current state."

On the other hand, typical computer science textbooks introduce algorithms in terms of *inputs* and *outputs* and then characterize algorithms as finite procedures (i.e., a finite set of instructions) for solving problems. Many books mention that an algorithm may or may not require input. An algorithm is said to imply instructions for a process that *creates* output [11]. The algorithm is unambiguous—that is, all steps of the procedure must be clear and well defined for the executor. Sedgewick and Wayne [12] state that an algorithm is used to describe a finite, deterministic, and effective problem-solving method. They define algorithm by describing a procedure for solving a problem in a natural language or by writing a computer program that implements the procedure. *Encyclopedia Britannica* [13] defines an algorithm as a purely mechanical procedure which determines the value of a function (e.g., truth-tables provide an algorithm for deciding whether any well-formed formula is tautologous).

According to Moschovakis [14], when algorithms are defined rigorously in the computer science literature, they are generally identified with abstract machines and mathematical models of computers, but this does not square with intuitions about algorithms and the way to interpret and apply results about them.

    Consider, for example, a function f : N → N on the natural numbers which is Turing computable, or, equivalently general recursive Consider, for example, a function f : N → N on the natural numbers which is Turing computable, or, equivalently general recursive. Now, there are many algorithms for computing f : the claim is that the "essential, implementation-independent properties" of each of them are captured by a recursive definition, while some "algorithm which compute f" cannot be "represented faithfully" by a Turing machine—or any other type of machine, for that matter. Moreover, this failure of expressiveness of machine models is even more significant for algorithms which operate on "abstract data" or "run forever", interacting with their environment. [14].

The variety of definitions for algorithms supports the argument that there is a need to devise better models (i.e., better ontological theories) of entities such as algorithms. This paper proposes diagrammatic definition of algorithms based on a new modeling machine called a thinging machine (TM). The machine has five actions (e.g., create, process, release, transfer, and receive) that form a network of flowing things (e.g., data). Thus, a thing can be created, processed, released, transferred, and received. TM creates, processes, releases, transfers, and receives things.

Before introducing the TM-based definition of an algorithm, the next section reviews the TM that has been developed by a series of published papers (see [15–24]).

## 2. Thinging Machine Modeling

The main TM thesis is that any entity has a double nature as (i) a thing and (ii) a process (abstract machine); thus, we call these things/machine entities *thimacs*. In TM modeling, intertwining with the world is accomplished by integrating these two modes of being of entities. Thimacs inhibit traditional categorization, properties, and behavior, replacing them with the five actions: creating, processing, releasing, transferring, and receiving. Such a thesis implies that all actions in a system can be reduced these to five generic (elementary) actions. Since generic events (to be defined later) are time-injected actions, there are five generic events. Because machines are things, all things can be reduced to five elementary things. These ideas were inspired by and can be traced back to Aristotle in ancient history and Heidegger in modern times (see [15–24]).

In TM modeling, a thing's machine operates on other things by creating, processing, releasing, transferring, and/or receiving them. The term "machine" refers to a special abstract machine (see Fig. 1). Among the five stages, flow (represented by the solid arrows in Fig. 1) signifies conceptual movement from one machine to another or among a machine's actions.

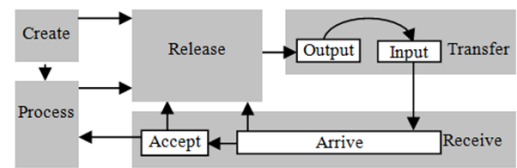

Fig. 1 Elementary machine

The TM's actions (also called stages) can be described as follows:

- *Arrival*: A thing reaches a new machine.
- *Acceptance*: A thing is permitted to enter the machine. If arriving things are always accepted, arrival and acceptance can then be combined into the "receive" stage. For simplicity, this paper's examples assume a receive stage exists.
- *Processing* (change): A thing undergoes a transformation that changes it without creating a new thing.
- *Release*: A thing is marked as ready to be transferred outside of the machine.
- *Transference*: A thing is transported somewhere outside of the machine.
- *Creation*: A new thing is born (is created/emerges) within a machine. A machine creates in the sense that it finds or originates a thing; it brings a thing into the system and then becomes aware of it. Creation can be designated "bringing into existence" in the system because what exists is what is found. Additionally, creation does not necessarily mean existence in the sense of being alive. Creation in a TM also means appearance in the system. Appearance here is not limited to form or solidity but also extends to any sense of the system's awareness of the new thing.

In addition, the TM model includes
- Memory and
- Triggering (represented as dashed arrows), or relations among the processes' stages (machines); for example, the process in Fig. 1 triggers the creation of a new thing.

To approach TM modeling smoothly, we focus on the machine side of thimacs. TM modeling is a three-level process that involves the following:

- A static model of the state of affairs to produce an atemporal diagrammatic description denoted as **S**. The state of affairs and actions are caused by the mixture of thimacs that penetrate each other part (e.g., process, receive). The time of S is present in the sense that everything subsists now.
- A decomposition of S into subdiagrams that forms the base of temporal events.
- The behavior of the model, denoted as B, formulated as a chronology of events. Behavior refers to executing composite actions.

## 3. Example of TM Modeling

TM modeling can be applied in a variety of systems to create a representation of a portion of reality.

In this section, we demonstrate TM modeling by re-describing an example that was developed using the data flow diagram (DFD). Thus, we not only show the general TM features but also draw a contrast between TM and a well-known DFD.

Karaca [25] gives an example of a DFD model of data flow through an information system. DFDs' main concern is with the movement of data among processes. Karaca's [25] example involves a restaurant that uses a system that takes customer orders, sends the orders to the kitchen, monitors goods sold and inventory, and generates reports for management.

Fig. 2 shows the TM static model of the example. In the figure does the following:

- The customer creates an order (circle 1) that flows to the system, (2) where it is processed, triggering
  - creation of a receipt (3) that flows to the customer (4);
  - transmission of the order to the kitchen (5 and 6);
  - creation of a record of the order (7) that flows to a machine (8) along with the current file (sold) (9), where they are processed (10) to produce a new version of the file (sold) (11); and
  - flow of the inventory files (12) along with the new record (13) to a machine, where they are processed (14) to create the new inventory file (15).
- The two files, inventory and sold, flow to a machine (16 and 17), where they are processed (18) to produce a report (19) that then flows to the manager (20).

### 3.1 Analysis of staticity.

The TM model of Fig. 2 is a static specification (S) that represents a rest (no time) condition. Additionally, it lacks the organization necessary for the flow of "information." This flow of information needs the notion of time, which is to be superimposed over S in the next phases of TM modeling. According to Wheeler's famous quote [26], "Time is Nature's way to keep everything from happening all at once." In S, everything is happening all at once, and contradiction is resolved when time is incorporated. Time is incorporated into the model through the notion of events. Events are defined over pieces, called *regions*, of the static description. Devising these regions requires decomposing Fig. 2. This analysis can be illustrated as follows: Decomposition→Regions→Events (in Time) →Behavior specification (chronology of events). That is, decomposition generates regions that, by applying time, create events. The system's behavior is specified through the chronology of the events.

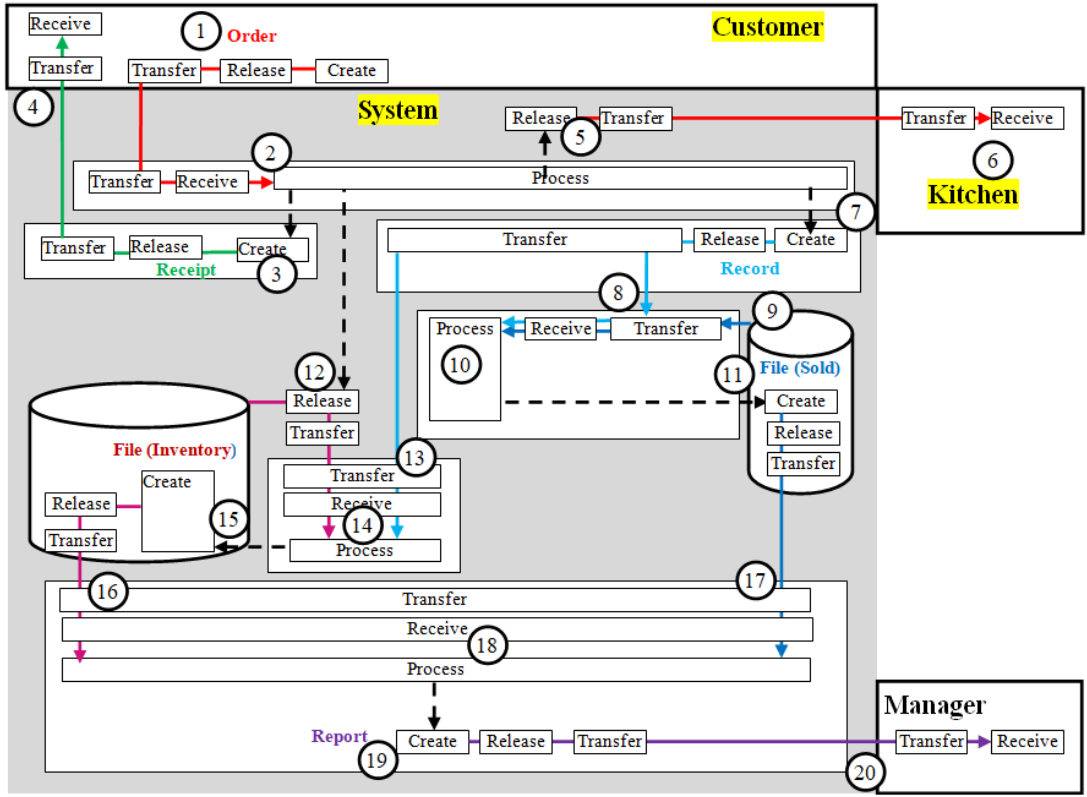

Fig. 2 The static TM model of the restaurant example

Forming an assemblage of he fragments of S is an evolvement of the initial whole to facilitate dynamism. Dynamism refers to temporal events. The selected subdiagrams of S become new thimacs, and the original thimac becomes a network of subthimacs. This is an evolutionary change from thimac S to its parts/subthimacs utilized to identify the abstract notion of the system's behavior.

## 3.2 Dynamics

Consider *The customer places an order to be processed by the system* as an event. Fig. 3 shows the TM model of the event. For simplicity's sake, we will represent an event by its region only. Accordingly, the dynamic model of the example (Fig. 4) involves the following selected events:

Event 1 ($E_1$): The customer places an order to be processed by the system.
Event 2 ($E_2$): A receipt is sent to the customer.
Event 3 ($E_3$): The order is sent to the kitchen.
Event 4 ($E_4$): A record is created and added to the file (Sold).

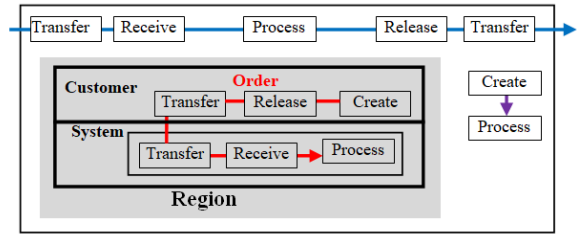

Fig. 3 The event *The customer places an order to be processed by the system.*

Event 5 ($E_5$): The record is subtracted from the inventory file.
Event 6 ($E_6$): The two files, inventory and sold files, are processed to produce a report that is sent to the manager.

These events are the unions of generic events.

Fig. 5 shows the behavioral model of the restaurant system. Note that an event never repeats; hence, in Fig. 5, the cycles are a notational simplification that indicates a sequence of events over those same regions.

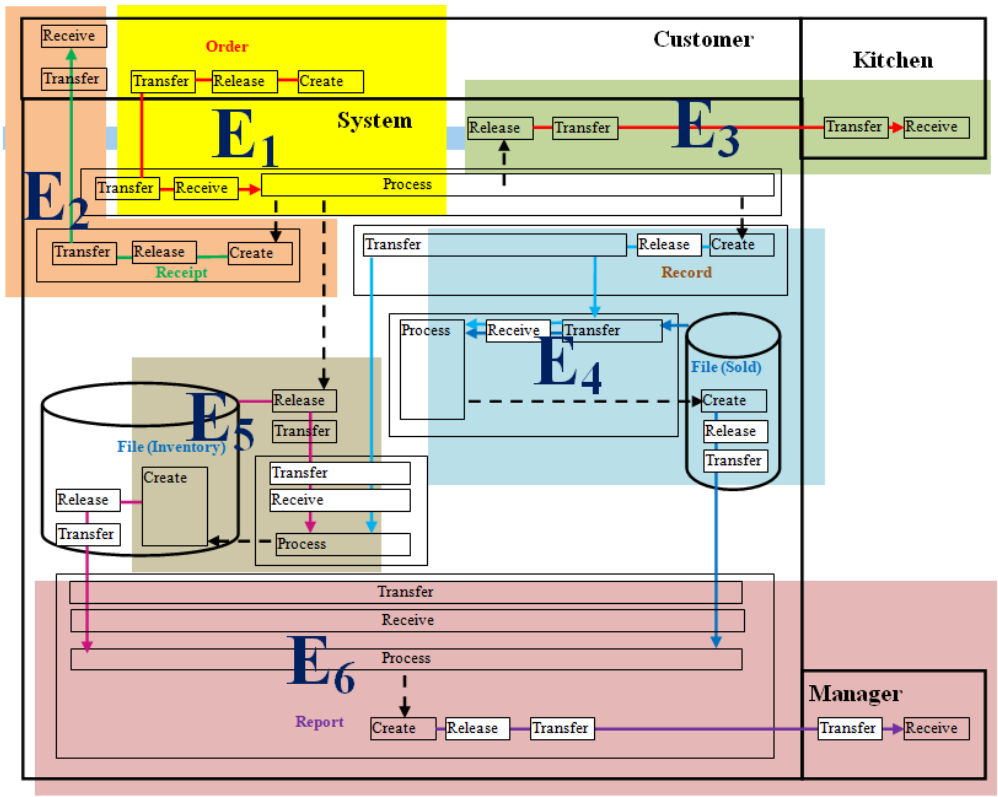

Fig. 4. The static TM model of the restaurant example.

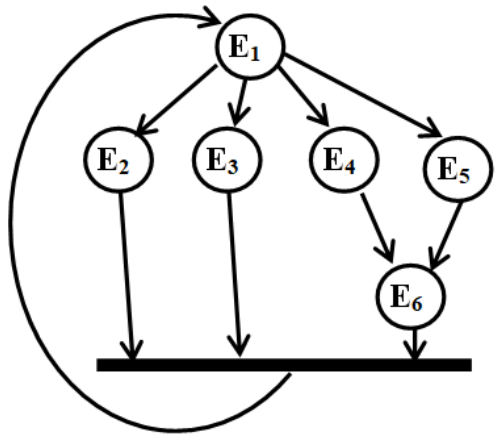

Fig. 5 The behavioral model of the restaurant system.

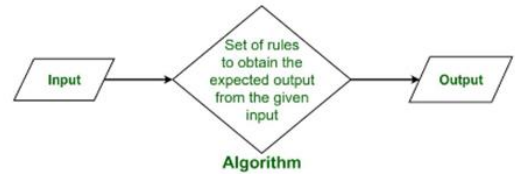

Fig. 6 What is algorithm (adapted from GeeksforGeeks [28]).

In a slide from a course at Carnegie Mellon University [27], an algorithm is described as follows: input→ algorithm→ output. We find this view (see Fig. 6) on many internet sites (e.g., [28]). As an analogy to such illustrations, Fig. 7 shows our basic description of an algorithm in terms of a single TM.

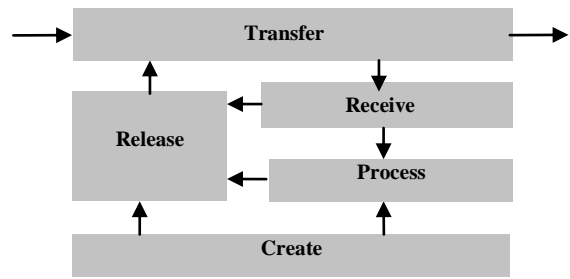

Fig. 7 A single machine algorithm.

In general, the TM algorithm is defined a network of TMs in which,

- each TM consists of the five elementary actions: *create*, *process*, *release*, *transfer*, and *receive* or subsets of these actions, where the arrows denote the flow of data;
- *Transfer* connects the machine to other machines through the sub-actions *input* and *output*; and
- *Receive* is the guard of the machine that includes the sub-actions *arrive* and *accept*.

Additionally, each machine may have

- Storage and
- Triggering, represented by a dash arrow, that indicates activation of other actions.

The behaviour of an algorithm is described by the chronology of finite time events defined over the network of TMs.

**Example 1**: Consider a single machine algorithm that receives the input of ten integers and outputs them. Fig. 8 shows the flowchart of the algorithm, and Fig 9 shows the static description, S, of the algorithm. It is assumed that all integers are accepted and that the direction of the arrow implies input/output in transfer. Fig. 10 shows a selection of events in S. Fig. 11 models the behavior, B, described in terms of two events:

Event 1: Inputting the integers
Event 2: Outputting the integers

We can choose to have one composite event that includes both events 1 and 2.

The flow chart (Fig. 8) mixes the static description with events. The dark part of the flow chart corresponds to the static TM model. The rest of the flowchart is the control of the loop. Dividing the definition of an algorithm into static and behavioral parts clarifies the difference between instructions and events. For example, we find in many programming courses the following statements:

- "Iteration is the term given to the repetition of a block of instructions (code) within a computer program for a number of instances or until a status is encountered." [The repetition refers to the *execution* of a block of instructions.]
- "Loop is defined as repeating a sequence of instructions a certain number of times." [The meaning is repeating the *execution* of a sequence of instructions.]

Separating the static description in an algorithm from its events eliminates such imprecision.

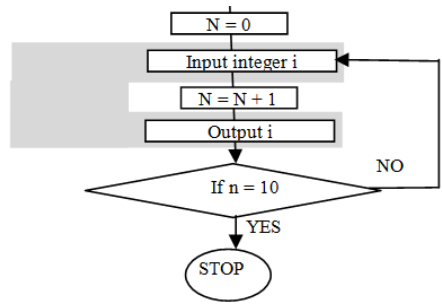

Fig. 8 The flowchart of the problem of example 1

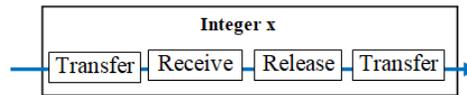

Fig. 9 An algorithm that receives integers and output them

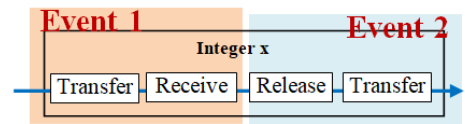

Fig. 10 An algorithm that receives integers and output them

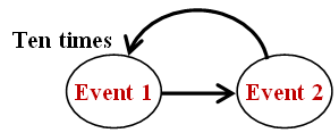

Fig. 11 The behavior of the algorithm that receives integers and output them

Note that the *iteration* is a second-level instruction (i.e., an instruction about instruction). An event never repeats; hence, Fig. 11 is a simplification that indicates a sequence of events over the same set of regions.

Fig. 9, as a static representation of an algorithm, involves *inputs* and *outputs* through the transfer stage. As mentioned before, the direction of the arrow implies input/output in transfer. Additionally, The representation reflects a *finite* procedure for solving the involved problem. Computer textbooks declare that *finiteness* means an algorithm should produce the output after *a finite number of* execution steps (TM events) for any input where the order of computation is critical to the functioning of the algorithm. Finiteness can be guaranteed by the absence of cycles in the TM diagram. This should not be confused with the simplification cycles discussed in the example.

Definiteness of events in a TM is a result of the genericity of the events. For example, we claim that there is no additional primitive event in the *create* event. Effectiveness (performing each step correctly and in a finite amount of time) can be described in terms of the generic events and their times in the behavioral model.

Admittedly, such claims about finiteness, definiteness and effectiveness need more verification. Nevertheless, the TM definition of an algorithm presents an alternative description that seems to be more precise than a natural language definition, and it seems to be closure to formal definitions such as the Turing machine-based definition. As a first attempt in that direction, the rest of this paper describes the application of TM modeling to algorithm-related topics.

**Example 2**: Figs. 12–14 show the static/events and behavioral representations of the algorithm that outputs "Odd" or "Even" depending on the input value.

## 5. Turing Machine

Consider how to design a Turing machine for the language 01*0 of length n. Fig. 15 shows a description of the machine as given by the Neso Academy [29]. Since this example involves many events, it is best to explain it in terms of its events and the behavioral model. Figs. 16 and 17, respectively, can be explained as follows.

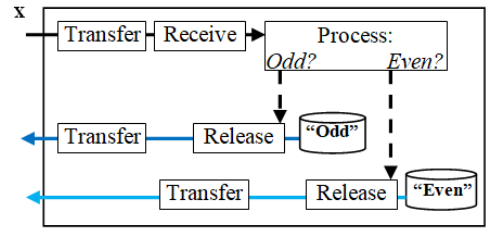

Fig. 12 The algorithm of outputting "Odd" of "Even"

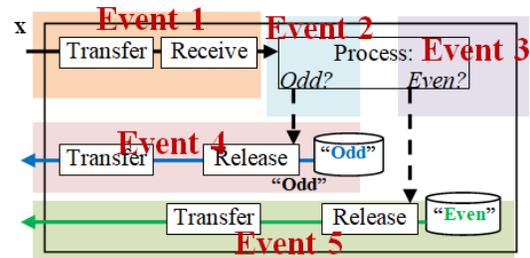

Fig. 13 Events of the algorithm of outputting "Odd" or "Even."

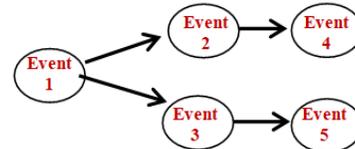

Fig. 14 The behavior of the algorithm of outputting "Odd" or "Even"

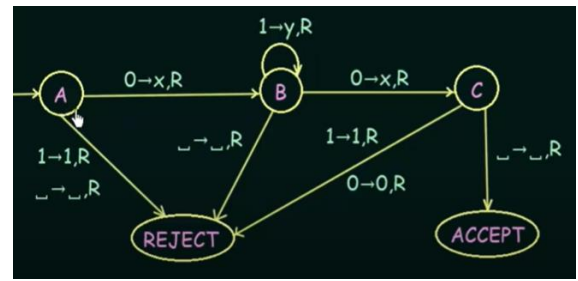

Fig. 15 A Turing machine that accepts 01*0 (adapted from [29]).

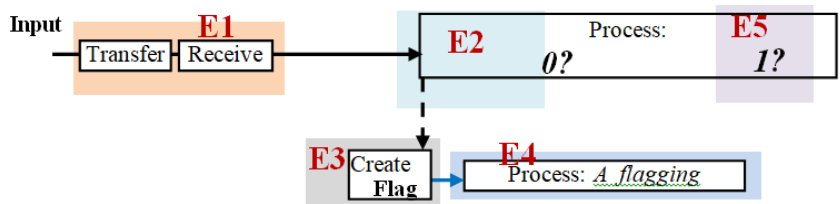

Fig. 16 TM model that accepts 01*0 with events

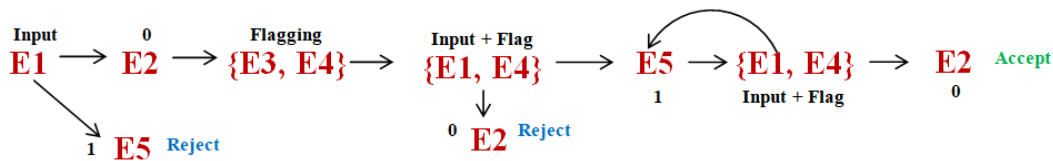

Fig. 17 The behavior of the TM model that accepts 01*0

In Fig. 16,

- The 0s or 1s are received (E1).
- If the first input is 1 (E5), then the string is rejected (see Fig. 17 – no more events).
- If the first input is 0 (E2), then a flag is created (3) and raised (4) – see Fig. 17.
- If the second input is 0 while flagging (E4), then the string is rejected. Note in this case that we have the composite event (E2 and E4). Note also that E4 is a continuing process of raising a flag.
- If the second input is 1 (E5) while flagging, then this is repeated until the input is 0. In this case, the string is accepted.

The result is an alternative representation that seems to be less abstract than Fig. 15.

## 6. Turing Machine That Checks the Palindrome

Consider how to design a Turing machine that checks the palindrome of the string of even length (from [30]). Fig. 18 shows a transition diagram of the machine given by Javapoint [30]. Fig. 19 shows the corresponding static and TM representation. In the figure, j (circle 1) and i (2) are the indices of the 0s and 1s in the given string, 01*0. We assume that i and j are initialized to 1 and n (the length), respectively. Each of the indices i and j, along with the string, are processed (3 and 4, respectively) to generate the corresponding value in the string (5 and 6, respectively).

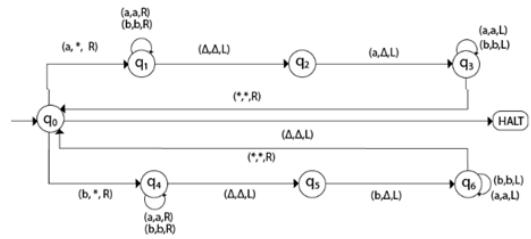

Fig. 18 The transition diagram of the Turing machine (from [30])

These values move (7) to be evaluated (8). According to the results of the comparison,

- if the two values are different, then the string is not palindrome;
- if the two values are equal, then the value i is incremented and updated (10 and 11), and the value of j is decremented by 1 (12); and
- if the result is less than n/2=1, there is a palindrome (13),
- otherwise, the value of j is updated (14).

Accordingly, the above process is repeated.

Fig. 20 shows the events of the model as follows:

$E_1$: The index j is processed along with the given string.
$E_2$: The index i is processed along with the given string.
$E_3$: The value in the string indexed by j is found.
$E_4$: The value in the string indexed by i is found.
$E_5$: The values indexed by i and j flow to be compared.

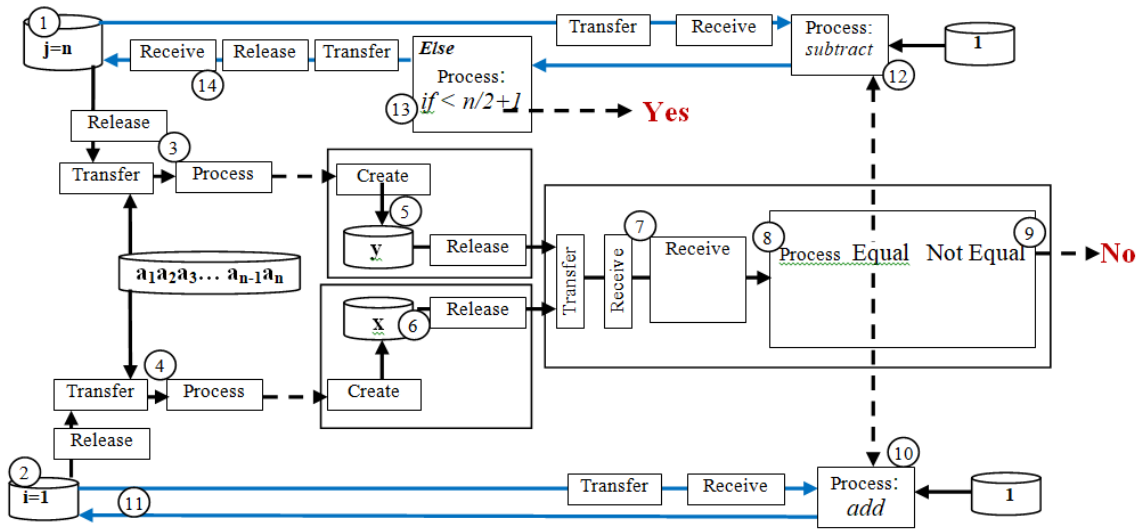

Fig. 19 The static TM representation of the Turing machine

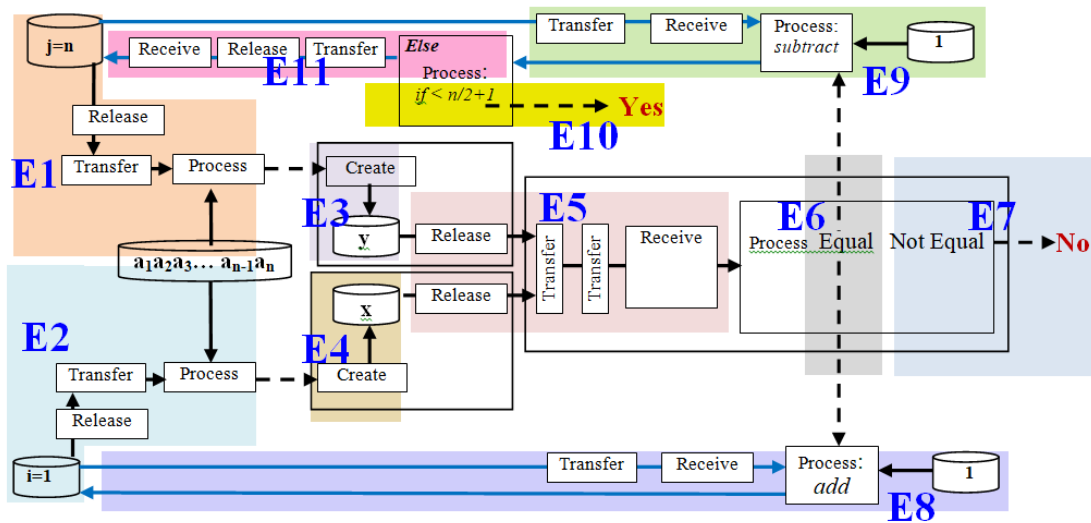

Fig. 20 Events in TM representation of the Turing machine

$E_6$: The values are equal.
$E_7$: The values are not equal; hence, there is no palindrome.
$E_8$: The index i is incremented by 1 and updated.
$E_9$: The index j is decrement by 1.
$E_{10}$: The index j is less than $(n/2)+1$; hence, there is a palindrome.
$E_{11}$: The index j is incremented by 1 and updated.

Fig. 21 shows the behavior of the model according to the chronology of events. It seems that the TM representations are less abstract than the transition diagram of the Turing machine with a reasonable level of preciseness. Further future exploration would clarify this point.

# 7. States and Discrete Event Semantics

In this section, we discuss a fine point in modeling that is related to the simultaneity of events, which illustrates how TM modeling clarifies such a notion in comparison with other models such as state machines. The issue is about transitions time in FSM.

Lee and Seshia [10] provide an example related to heating, ventilation, and air-conditioning systems, since an "accurate model of temperature dynamics and temperature control systems is a significant feature of the system." The modeling begins with a thermostat, which regulates temperature to maintain a target temperature.

Consider a thermostat modeled by an FSM with States = {heating, cooling} as shown in Figure [22]. Suppose the setpoint is 20 degrees Celsius.

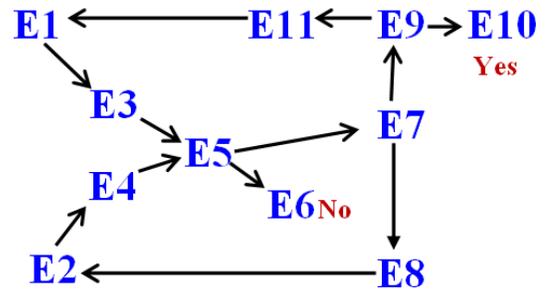

Fig. 21 The behavior of TM model of the Turing machine

If the heater is on, then the thermostat allows the temperature to rise past the setpoint to 22 degrees. If the heater is off, then it allows the temperature to drop past the setpoint to 18 degrees. This strategy is called hysteresis. It avoids chattering, where the heater would turn on and off rapidly when the temperature is close to the setpoint temperature.

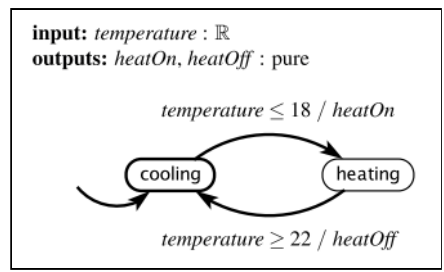

Fig. 22 Model of the thermostat (Adapted from Lee and Seshia [10]).

There is a single input temperature with type R and two pure outputs heatOn and heatOff. These outputs will be present only when a change in the status of the heater is needed (i.e., when it is on and needs to be turned off, or when it is off and needs to be turned on) (Lee and Seshia [10]).

Fig. 23 shows the corresponding TM model. To save space, we show only the events diagram and the behavioral model (Fig. 24). In Fig. 23, the outside environment triggers creation of the temperature degree in the thermostat (circle 1). The temperature value is processed according to the following.

- If the temperature $\leq$ 18 degrees (2), then cooling is turned OFF if it is ON (3), and the heating is turned ON if it is OFF (4). Note that such an arrangement in the cooling and heating is made to avoid continuously turning the boundary degrees ON and OFF. For example, heating is turned ON when the signal of $\leq$ 18 degrees is received (triggering) the first time. But if, after the first turning ON, a signal of a temperature close to 18 continues to be received (called chattering by Lee and Seshia [10]), the heating ignores such a signal.
- If the temperature $\geq$ 22 degrees, then the heating is turned OFF if it is ON (3), and the cooling is turned ON if it is OFF (5).

Fig. 24 shows the behavioral model, B. It is not difficult to specify the events in the figure as we did in the previous modeling examples. Event $E_1$ in Fig. 23 continuously happens and thus is followed by other events. As mentioned previously, the reflexive arrow is a convenient notation to indicate events on the same region.

Lee and Seshia [10] define the *state* of a system as "its condition at a particular point in time." In general, the state affects how the system reacts to inputs. Formally, Lee and Seshia [10] define the state as an encoding of everything about the past that has an effect on the system's reaction to current or future inputs.

It can be noted that the condition of the whole system is related to the synchronization of the cooling and heating subsystems. The decomposition of the system in a certain way provides more elementary states. The cooling or heating state in the thermostat system includes two substates. The events that are based on the decomposing can be grounded in the TM's generic actions.

Lee and Seshia's [10] model does not mention a clear definition of events. In one example in their work, they claim that a detector (of cars) actor produces events. Also, a counter actor produces an output event that updates a display. A discrete event occurs at an instance of time rather than over time. Discrete signals consist of a sequence of instantaneous events in time. According to Lee and Seshia [10], the actions on one or more state transitions define the discrete *event behavior* that combines with *time-based behavior*. Discrete events (*state changes* in a state machine) become embedded in a time base.

In general, a discrete event is typically described as something that occurs instantaneously that may change the state of the system. In a TM, the generic events that correspond to the generic actions (e.g., create, process) are taken as the minimum existence of an event. Switching from a generic event (e.g., create) to another (e.g., process) occurs spontaneously. For example, a typical instantaneous event is the arrival of a customer (in simulation). In a TM, this event is expressed in terms of transfer (output), transfer (input in queue), and receive. In this case, transfer to transfer is instantaneous. It is a generic change that cannot be discomposed by a further change. In a TM, instantaneity is fixed to generic events.

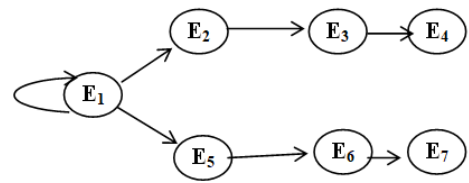

Fig. 24 The behavioral model of the thermostat

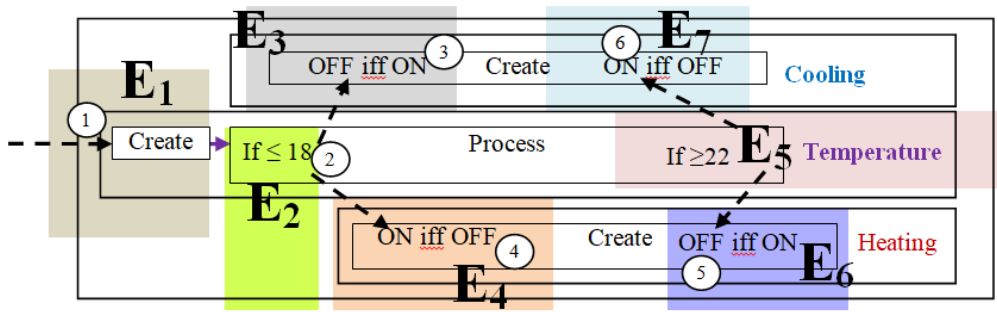

Fig. 23 Events in the TM model of the thermostat.

According to Marwedel [31], edges in FSMs denote state transitions, and edge labels represent events. Hence, *create temperature measure* in a thermostat state machine is an event. Internal events are generated as a result of some transition and are described in reaction-parts. As mentioned, events are generated either internally or externally. External events are usually described in the model environment. If an event happens, the FSM changes its state as indicated by the corresponding edge, according to Marwedel [31].

In TM, the event that creates *temperature measure* is a series of the generic events (process, release, etc.). The change between generic events (e.g., create to release) is instantaneous. Thus, if we have $E_1$ and $E_2$, then one event just happened and another is about to happen, but neither is ever that which is happening (these ideas are from Deleuze [32]). Note that, in a TM, an event never repeats and the modeling of a repeated event (reflexive arrow in a machine) in TM diagrams is a simplification of a set of events within the same region (subdiagram).

Accordingly, the notion of transition (as in FSM) is applied to generic events in TM. The transition among generic events is instantaneous in the sense that the *between being* is eluded. This is analogous to the supposition that there is no integer between 1 and 2. To explore this difference in the notion of an event, we consider examples from discrete event semantics.
.

According to Vangheluwe [33], "Untimed formalisms such as State Automata, State Charts, and Petri Nets were introduced. In these formalisms, only the order in which events occur is represented, not the explicit time at which they occur." This implies a loss of information. State automata can be extended to include the time the system stays in a particular state before making a transition to the next state [33].

As shown in Fig. 25, it is possible to construct an event graph that has the transitions as nodes and, as edges, the time interval after which the next transition is scheduled to occur. This demonstrates the link with the scheduling of a discrete event worldview introduced in the next section [33].

The point here is that FSM instantaneous events reflect processes that take time, just like in the case of states. This raises questions about the benefit of distinguishing states from transitions. Fig. 26 shows the TM model of the red-green-yellow traffic system. Accordingly, the events are specified as follows:

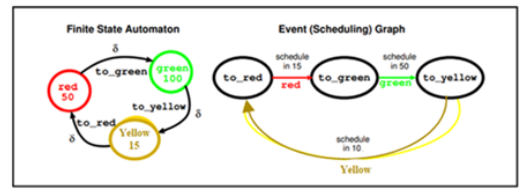

Fig. 25 FSM and event graph models (From [33])

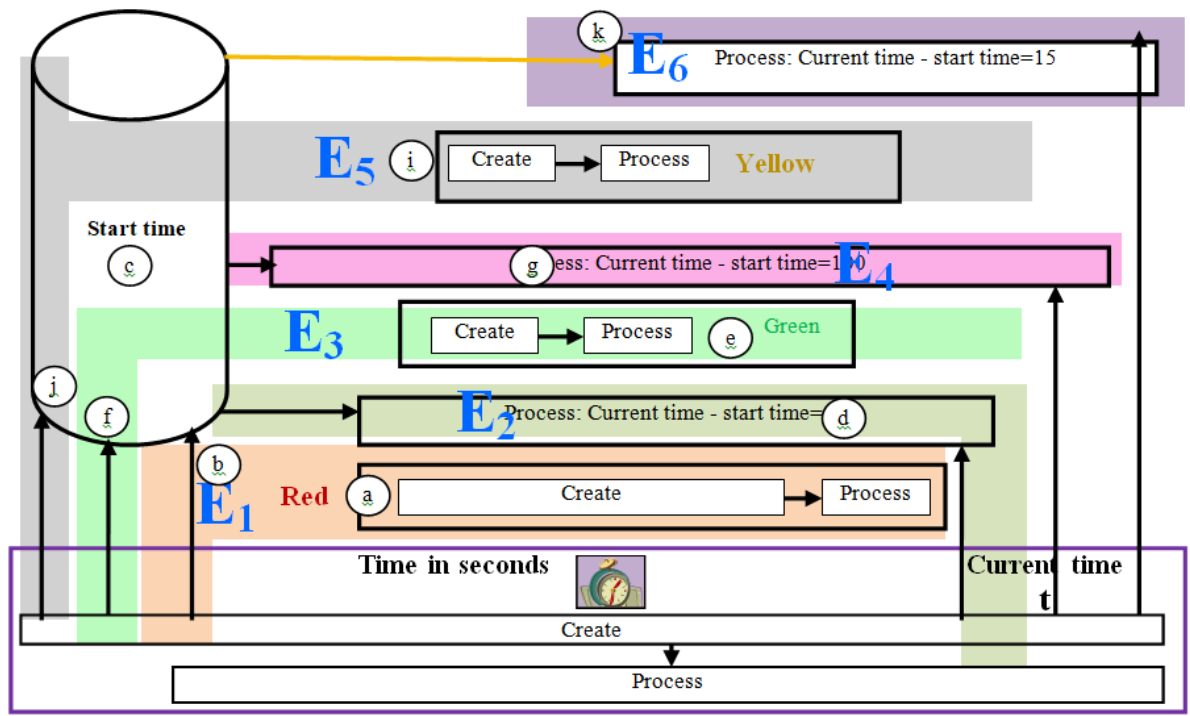

Fig. 26 FSM and event graph models (From [33])

Event 1 ($E_1$): The light is initially red (circle a - assumption), and the starting time (b) of this event is stored in *start time* (c).

Event 2 ($E_2$): The time of the red light (d) is calculated: (t-c) = 50.

Event 3 ($E_3$): The light turns green (d), and its starting time is recorded (f) in *start time* (c).

Event 4 ($E_4$): The time of the green light (g) is calculated: (t-c) = 100.

Event 5 ($E_5$): The light turns yellow (i), and its starting time is recorded (j) in *start time* (c).

Event 6 ($E_6$): The time of the yellow light (k) is calculated: (t-c) = 15.

Fig. 27 shows the behavioral model, including the time periods of events. The event $E_1$ event lasts 50 seconds. The calculation ((t-c) = 50) consumes time in determining the next event. $E_2$ is called transition in FSM, but in TM, it is part of a composite event that should occur before the end of 50 seconds (see Fig. 28). E2 joins E1 to form a composite event. Since the time of $E_2$ is less than $E_1$, generating such synchronization is possible. The assumption is that $E_2$ would use its time to determine the event in the future that will follow $E_1$. The point here is that there is no need to distinguish states from events in a TM, non-generic transition.

## 8. Conclusion

This paper proposes a diagrammatical definition of an algorithm based on a new modeling machine called a thinging machine. The proposed definition appears to have a wider implication that obscures the boundaries between algorithmics and modeling. The paper explores applying the definition in Turing machines and FSMs. This research is a first attempt at diagrammatic modeling of algorithms intended to furnish a more precise description of the commonly used informal definition (e.g., set of instructions); however, the results indicate that the proposed approach can be applied to issues in algorithm-related fields of study.

## References

[1] Sanders, P., Wagner, D., Karlsruhe Institute of Technology: *Algorithm Engineering*. Information Technology, 53(6), 263–265 (2011). DOI: 10.1524/itit.2011.9072

[2] Kliemann, L., Sanders, P. (Eds.): *Algorithm Engineering Selected Results and Surveys*. Theoretical Computer Science and General Issues, Series vol. 9220. Springer International Publishing, 2016. DOI: 10.1007/978-3-319-49487-6

[3] Kolmogorov, N.A.: *On the Concept of Algorithm*. Uspekhi Mat. Nauk 8(4), 175–176 (1953)

[4] Gurevich, Y.: *What Is an Algorithm*? In: Bielikova, M., et al. (eds.) SOFSEM 2012. Theory and Practice of Computer Science. LNCS, vol. 7147, pp. Springer, Berlin (2012)

[5] Rapaport, W.J.: *Philosophy of Computer Science: An Introductory Course*. Teaching Philosophy 28(4), 319–341 (2005)

[6] Kleinberg, J.: *The Mathematics of Algorithm Design, 2008*, OAI identifier: oai:CiteSeerX.psu:10.1.1.62.603. Accessed Nov. 3, 2020.
https://www.cs.cornell.edu/home/kleinber/pcm.pdf

[7] Miłkowski, M.: *Computational Mechanisms and Models of Computation*. Philosophia Scientiae 18(3), 215–228 (2014). DOI: 10.4000/philosophiascientiae.1019

[8] De Mol, L.: *Turing Machines*. In: Zalta, E.N. (ed.) The Stanford Encyclopedia of Philosophy (2019). https://plato.stanford.edu/archives/win2019/entries/turing-machine/

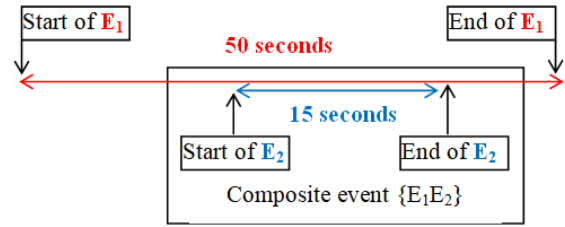

Fig. 28 The relation between the composite event and $E_1$.

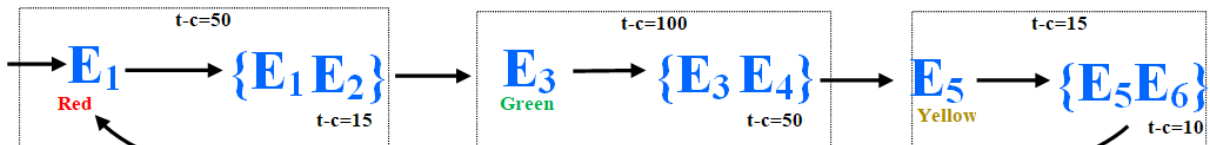

Fig. 27 The behavioral model

[9] Bendersky, E.: *Finite State Machines and Regular Expressions*. Gamedev.com, June 12, 2013. https://www.gamedev.net/tutorials/programming/general-and-gameplay-programming/finite-state-machines-and-regular-expressions-r3176/

[10] Lee, E.A., Seshia, S.A.: *Introduction to Embedded Systems*. In: A Cyber-Physical Systems Approach, Second Edition. MIT Press, 2017.

[11] Boolos G., Jeffrey R.: Computability and Logic, First Edition. Cambridge University Press, London (1974). https://psychology.wikia.org/wiki/Algorithms

[12] Sedgewick, R., Wayne, K.: Algorithms, Fourth Edition. Pearson Education, Inc., Boston (2011)

[13] Britannica Encyclopedia. *Algorithm*. Accessed Nov. 3, 2020. http://www.philosophypages.com/dy/a2.htm

[14] Moschovakis, Y.N.: *What Is an Algorithm?* In: Engquist, B., Schmid W. (eds.) Mathematics Unlimited — 2001 and Beyond. Springer, Berlin, Heidelberg (2001). https://doi.org/10.1007/978-3-642-56478-9_46

[15] Al-Fedaghi, S., Al-Fadhli, J.: *Thinging-Oriented Modeling of Unmanned Aerial Vehicles*. International Journal of Advanced Computer Science and Applications 11(5), 610–619 (2020). DOI: 10.14569/IJACSA.2020.0110575

[16] Al-Fedaghi, S., Behbehani, B.: *How to Document Computer Networks*. Journal of Computer Science 16(6), 423–434 (2020). DOI: 10.3844/jcssp.2020.723.434

[17] Al-Fedaghi, S., Al-Qemlas, D.: *Modeling Network Architecture: A Cloud Case Study*. International Journal of Computer Science and Network Security 20(3), 195–209 (2020)

[18] Al-Fedaghi, S., Bayoumi, M.: Modeling Advanced Persistent Threats: A Case Study of APT38. In: 14th International Conference for Internet Technology and Secured Transactions (ICITST-2019), London (2019)

[19] Al-Fedaghi, S., Aldamkhi, G.: Conceptual Modeling of an IP Phone Communication System: A Case Study. In: 18th Annual Wireless Telecommunications Symposium (WTS 2019), New York City (2019)

[20] Al-Fedaghi, S., Haidar, E.: *Programming Is Diagramming Is Programming*. Journal of Software, 14, 410–422 (2019). DOI: 10.17706/jsw.14.9.410-422

[21] Al-Fedaghi, S., Al-Otaibi, M.: Service-Oriented Systems as a Thinging Machine: A Case Study of Customer Relationship Management. In: Proceedings of the IEEE International Conference on Information and Computer Technologies, University of Hawaii, Maui College, Kahului, Hawaii, pp. 243–254 (2019). DOI: 10.1109/INFOCT.2019.8710891

[22] Al-Fedaghi, S., Makdessi, M.: Modeling Business Process and Events. In: 9th Computer Science On-line Conference, Springer, Applied Informatics and Cybernetics in Intelligent Systems, pp. 83–97 (2020). doi.org/10.1007/978-3-030-30329-7_8

[23] Al-Fedaghi, S., Haidar, E.: Thinging-based Conceptual Modeling: Case Study of a Tendering System. Journal of Computer Science 16(4) 452–466 (2020). DOI: 10.3844/jcssp.2020.452.466

[24] Al-Fedaghi, S., Al-Saraf, M.: Thinging the Robotic Architectural Structure. In: The 3rd International Conference on Mechatronics, Control and Robotics, Tokyo, Japan (2020)

[25] Karaca, K.: *Philosophical Reflections on Diagram Models and Diagrammatic Representation*. Journal of Experimental & Theoretical Artificial Intelligence 24(3), 365–384 (2012). DOI: 10.1080/0952813X.2012.693665

[26] Toggl Track: *50 Handpicked Time Quotes from Notable Thinkers Throughout History*. Accessed Oct. 10, 2020. https://toggl.com/blog/time-quotes

[27] Carnegie Mellon University, School of Computer Science: *Algorithmic Thinking: Loops and Conditionals, Principle of Computing Course Slides*. https://www.cs.cmu.edu/~15110-n15/lectures/unit03-Algorithm-1.pdf

[28] GeeksforGeeks: *Introduction to Algorithms, 16-10-2019*. Accessed Nov. 10, 2020. https://media.geeksforgeeks.org/wp-content/cdn-uploads/20191016135223/What-is-Algorithm_.jpg

[29] Neso Academy: *Turing Machine (Example 1), Presentation*. https://www.youtube.com/watch?v=D9eF_B8URnw

[30] Javapoint: *Examples of TM*. Accessed Oct. 7, 2020. https://www.javatpoint.com/examples-of-turing-machine

[31] Marwedel, P.: Embedded System Design. Springer, Dordrecht, The Netherlands (2006)

[32] Deleuze, G.: *Logique du sens* [The Logic of Sense]. Translated by Lester, M., Stivale, C. Columbia University Press, New York City (1990)

[33] Vangheluwe, H.: *Discrete Event Modelling and Simulation* [Lecture notes]. Accessed Nov. 1, 2020. https://www.cs.mcgill.ca/~hv/classes/MS/discreteEvent.pdf